\documentclass[a4paper,11pt]{article}

\usepackage{pos}
\usepackage{lipsum} 
\usepackage{gensymb}

\title{Reconstruction of inclined cosmic-ray properties with GRAND data}

\ShortTitle{Reconstruction of inclined cosmic rays on GRAND}

\author*[a, b]{Marion Guelfand}
\author[c]{Pauline Fritsch}
\author[d]{Valentin Decoene}
\author[a, b, e]{Olivier Martineau-Huynh}
\author[f]{Mauricio Bustamante}

\onbehalf{for the GRAND Collaboration{\normalsize \\ \normalfont(a complete list of authors can be found at the end of the proceedings)}\\}

\affiliation[a]{Sorbonne Université, CNRS, Laboratoire de Physique Nucléaire et de Hautes Energies (LPNHE), Paris, France}
\affiliation[b]{Institut d’Astrophysique de Paris, CNRS, Sorbonne Universite, Paris}
\affiliation[c]{ILANCE, CNRS – University of Tokyo International Research Laboratory, Kashiwa, Chiba 277-8}
\affiliation[d]{SUBATECH, Institut Mines-Telecom Atlantique, CNRS/IN2P3, Universite de Nantes, Nantes}
\affiliation[e]{National Astronomical Observatories, Chinese Academy of Sciences, Beijing, China}
\affiliation[f]{Niels Bohr International Academy, Niels Bohr Institute, University of Ceopenhagen, Denmark}

\emailAdd{marion.guelfand@lpnhe.in2p3.fr}

\abstract{
Radio-detection is now an established technique for studying ultra-high-energy (UHE) cosmic rays with energies exceeding $\sim 10^{17}$ eV. The next generation of radio experiments, such as the Giant Radio Array for Neutrino Detection (GRAND), aims to expand this technique to the observation of Earth-skimming UHE neutrinos, which requires the detection of very inclined extensive air showers (EAS). Currently, GRAND is validating its detection principle —autonomous radio detection —in particular through the prototype array GRANDProto300, deployed in the Gobi Desert. In this phase, the array is limited to detecting inclined EAS from cosmic rays. Neutrinos cannot be observed because of the restricted detector size. 
We present a method to reconstruct the arrival direction and energy of EAS with zenith angles above $60^\circ$, applicable as well to upward-going trajectories. The approach combines a point-source-like description of the radio wavefront with the so-called Angular Distribution Function (ADF), a phenomenological model describing the angular pattern of radio signal amplitudes in the 50–200\,MHz band. Applied directly to the voltage traces, this method enables efficient event selection with accurate direction reconstruction and a first-order energy estimate. We validate the approach with both simulations and experimental data, and reconstruct the first cosmic-ray candidates detected by GRANDProto300.

\vspace{4mm}

}

\ConferenceLogo{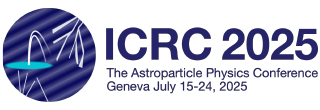}

\FullConference{39th International Cosmic Ray Conference (ICRC2025)\\  15--24 July, 2025\\ Geneva, Switzerland}

\begin{document}

\maketitle

\vspace{-0.3cm}
\section{Introduction}
\vspace{-0.3cm}
Radio detection has become a promising technique for detecting ultra-high-energy cosmic particles thanks to its continuous operation, robustness and cost-effective scalability over large areas. Next-generation experiments focus on very inclined air showers, which produce large radio footprints, enabling the deployment of sparsely spaced radio antennas to cover vast effective areas and probe very low fluxes. 

GRANDProto300 (GP300), one of the prototype arrays of the GRAND project \cite{Martineau_ICRC2025}, is currently under commissioning. It offers a unique opportunity to develop and test reconstruction methods directly with the experimental data. In this context, we explore the application of the ADF (Angular Distribution Function) model ~\cite{Guelfand_2025} directly on voltage traces, as recorded by the detectors, bypassing the full electric field reconstruction—although the model was originally developed for it.

First, we briefly review our reconstruction procedure and the ADF model. Then, we present the application of the ADF method to both simulated and experimental GP300 data, consisting of 41 selected cosmic-ray candidates. We demonstrate that applying the fit directly to voltage signals enables efficient event selection, accurate direction reconstruction, and a first-order estimate of the shower energy. A cross-check with the electric field-based reconstruction confirms the consistency and reliability of this voltage-based approach.

\vspace{-0.3cm}
\section{Reconstruction pipeline}
\label{sec:recons}
\vspace{-0.3cm}
The reconstruction procedure is based on the combined use of signal arrival times and peak amplitudes of the electric field at individual antennas. The peak amplitude is defined as the maximum of the Hilbert envelope of the 
electric field, computed as the modulus of its three polarization components (East-West, North-South, and Vertical). Its time marks the wavefront arrival.  The reconstruction proceeds in three steps, described in \cite{Decoene_2021, Guelfand_2025} and summarized below:

\vspace{-0.2cm}
\begin{enumerate}
    \itemsep0em
    \item \textbf{Initial direction estimate:}  
A plane-wave fit to the arrival times yields a first estimate of the shower direction ($\theta_0, \phi_0$) using the analytical approach of \cite{Ferriere_2025}. While optional, this step reduces the parameter space and improves convergence in subsequent steps.
    \item \textbf{Emission point reconstruction:}
The wavefront is then fitted with a spherical curvature model, relevant for very inclined showers \cite{Decoene_2023}, to determine the emission point $X_e$. The fit scans a 3D volume around the initial direction, accounting for atmospheric refraction along each antenna's optical path. The resulting $X_e$, combined with ($\theta_0, \phi_0$), defines an initial shower axis. 

    \item \textbf{Amplitude profile fit (ADF):} 
With $X_e$ fixed, the shower direction is refined by fitting the angular distribution of amplitudes in the shower plane using the Angular Distribution Function (ADF). The ADF fit minimizes: $R(\theta, \phi) = \sum_{i=1}^{N_{\rm ant}} \left[ A_i - f_i^{\rm ADF}(A, \theta, \phi, \delta \omega; X_e) \right]^2$,
\vspace{-0.1cm}
where $A_i$ is the measured peak amplitude and $f_i^{\rm ADF}$ models the expected amplitude as:
\vspace{-0.1cm}
\begin{equation}
\label{eq:adf}
f^{\rm ADF} (A, \theta, \phi, \delta \omega; X_e) = \frac{A}{l} \left[ 1 + G_A \frac{\cos \eta}{\sin \alpha} \right] \cdot \frac{1}{1 + 4 \left( \frac{(\tan \omega / \tan \omega_c)^2 - 1}{\delta \omega} \right)^2 } \ ,
\end{equation}
\end{enumerate}
\vspace{-0.1cm}
$(\omega, \eta)$ are angular coordinates of antennas in the shower plane relative to $X_e$, with $\omega$ the angular distance to the shower axis and $\eta$ the azimuth relative to $\mathbf{k} \times \mathbf{B}$, where $\textbf{k}$ is the propagation vector and $\textbf{B}$ the Earth’s magnetic field. The Cherenkov angle $\omega_c$ is computed from $X_e$ and refractive index along each antenna’s line of sight \cite{Guelfand_2025}. The geomagnetic asymmetry depends on the geomagnetic angle $\alpha$ and asymmetry strength $G_A$. The early-late effect is controlled by the geometric distance $l$ which accounts for antenna distance to $X_e$.

The fit adjusts four parameters: the shower direction ($\theta$, $\phi$), a global amplitude scaling $A$, and the width of the amplitude profile $\delta \omega$, which reflects the lateral spread of the radio signal around the shower axis. 
Although only $A$ and $\delta \omega$ appear explicitly in the ADF, the antenna angles $\omega$ and $\eta$ depend on ($\theta$, $\phi$) and $X_e$.

An additional step is required to estimate the electromagnetic energy of the shower. The key quantity is the scaling factor $A$ from the ADF fit (Eq.~\ref{eq:adf}), which directly reflects the amplitude of the radio signal. To obtain an energy estimator from $A$, we apply the following correction factors: i) a $\sin(\alpha)$ term, accounting for the impact of the shower geometry on the efficiency of the geomagnetic emission; ii) a factor $f(\rho, \sin\alpha)$, which primarily depends on the air density $\rho$ at the emission point and includes a secondary dependence on  $\sin(\alpha)$, capturing coherence effects \cite{Chiche_2024}. This factor is determined from simulations.

Our energy estimator can thus be written as\,: $E^{*}_{\rm em} = \frac{A}{\sin(\alpha) \, f(\rho, \sin(\alpha))}$.
\vspace{-0.3cm}
\section{Testing the ADF fit to Voltage Signals}
\vspace{-0.3cm}
\subsection{Motivations}
\label{subsec:motivations}
\vspace{-0.2cm}
The ADF model is initially tailored to be applied to the electric field to reconstruct the direction and energy of cosmic rays. Since detectors directly measure voltage signals, we investigate applying the ADF fit directly to these voltages. This approach not only enables event identification directly from raw data but also bypasses the electric field reconstruction step.

This approach is motivated by the characteristics of the GRAND Horizon antennas \cite{Alvarez_Muniz_2019}, which feature a wide and relatively flat gain at zenith angles $70^\circ$--$85^\circ$. In this regime, the EAS source is distant and the viewing angles across the array are thus nearly uniform. As a result, the antenna response varies little from one antenna to another, and the voltage is expected to scale approximately linearly with the electric field. It has been checked with simulations presented in the following section, by comparing the amplitudes of electric field traces with those of ADC traces. This should make a direct application of the ADF model to voltage data plausible. 

Finally, timing differences between voltage and electric field signals are negligible aside from a global offset, preserving arrival time reconstruction accuracy.
\vspace{-0.2cm}
\subsection{Reconstruction performances on simulations}
\vspace{-0.2cm}
\subsubsection{GP300 layout and simulation set}
\label{subsec:recons_simus}
\vspace{-0.1cm}

To validate applying ADF to voltages, a dataset of 13,000 air showers (proton and iron primaries, 0.1–3.98 EeV) was simulated with ZHAireS. Arrival directions are uniform in azimuth and follow $-\log(\cos\theta)$ in zenith ($60^\circ$–$88^\circ$), with cores randomly distributed in the planned GP300 layout: a hexagonal array of ~150 antennas spaced by 1 km, with a denser infill of ~134 antennas at 250 m spacing. The simulated electric field traces are passed through a modeled electronics chain to generate realistic voltage signals \cite{Alves_Batista_2025} and are filtered in the 50–200~MHz band, the nominal operating frequency range of the GRAND experiment.

To model the realistic conditions, the simulated voltage traces are superimposed with measured on-site stationnary background noise, 
with an additional Gaussian jitter of $\sim 5 $ns added to the GPS timing. An event is considered triggered if at least five antennas record peak amplitudes exceeding 100 ADC counts ($\sim 6\sigma$ above noise), a condition met by roughly 3 000 events.
\vspace{-0.1cm}
\subsubsection{Direction and energy reconstruction}
\vspace{-0.1cm}
The distribution of reduced $\chi^2_\nu = \chi^2/\rm{ndf}$ values obtained from the ADF fit on simulations is shown in orange in Figure~\ref{fig:voltage_CR0} {\it Left}. In the following, we select only events with $\chi^2_\nu \leq 25$, which keeps 86\,\% of the triggered events. The relatively high $\chi^2_\nu$ values likely result from the model being originally designed for electric field, not voltage.

\begin{figure}[htbp]
   \centering
      \centering
      \includegraphics[width=0.44\linewidth]{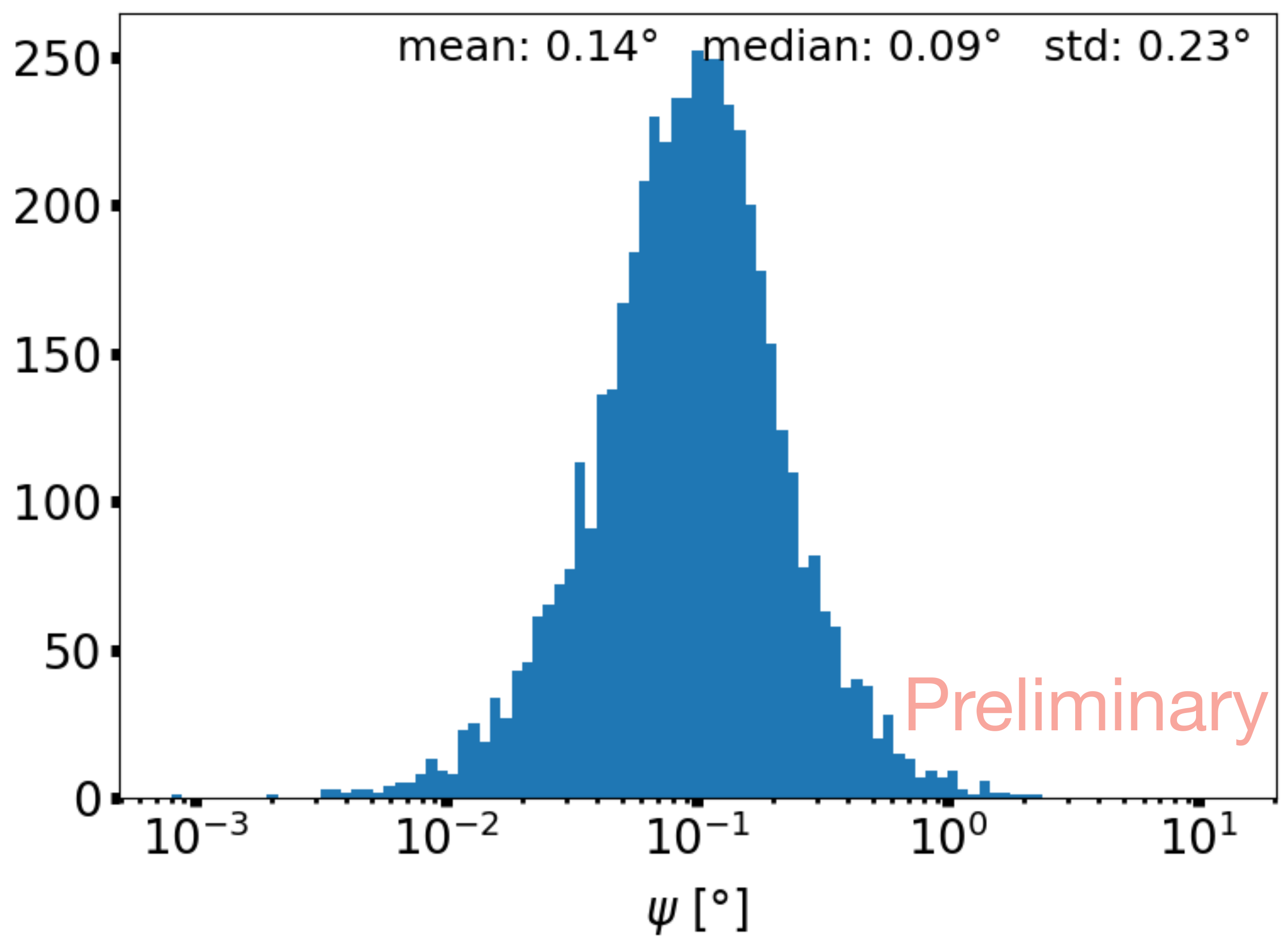}
      \includegraphics[width=0.44\linewidth]{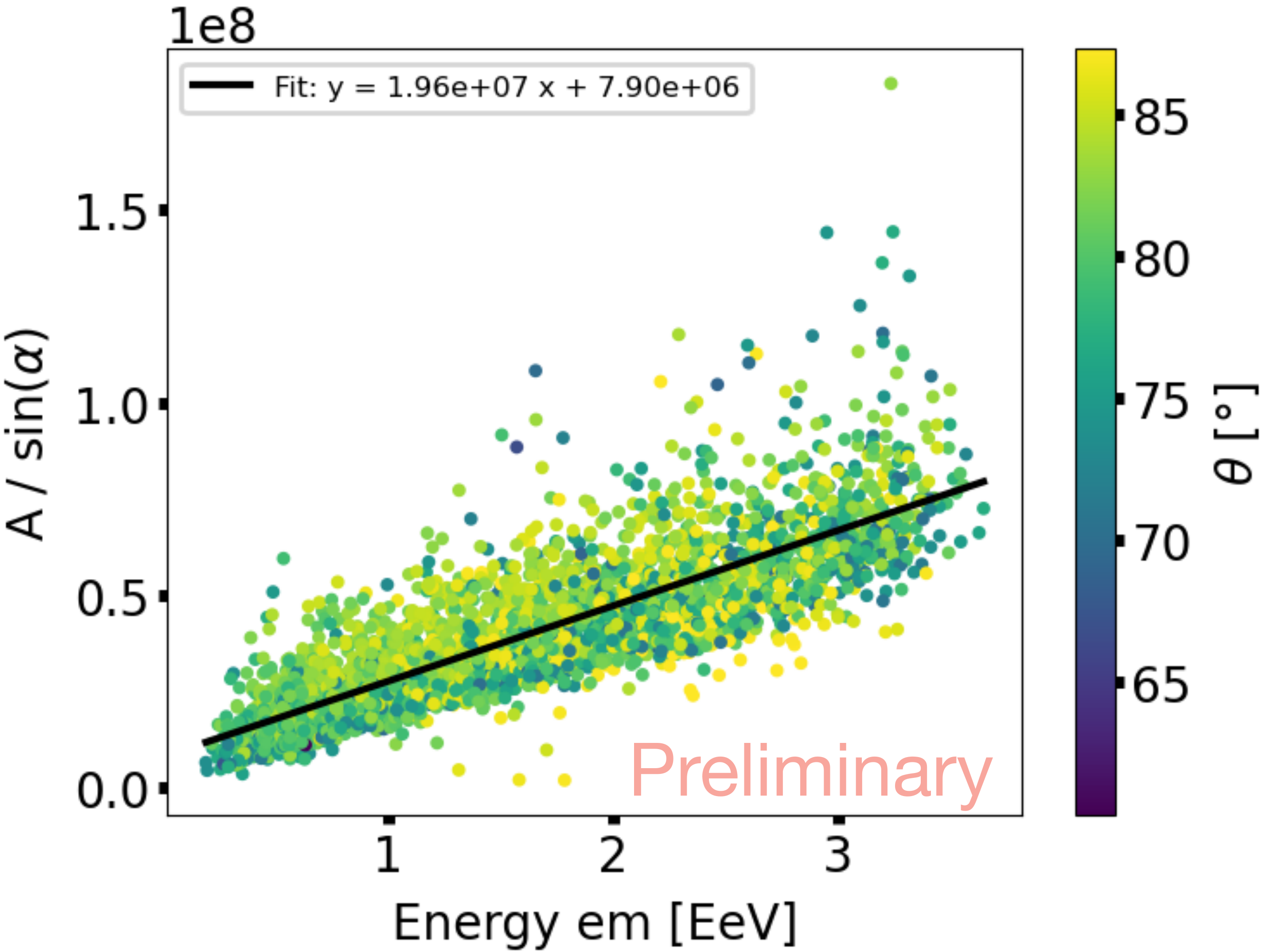}
  \caption{
   \textit{Left:} Angular resolution $\psi$ obtained directly from voltage using the ADF fit on simulations. 
   \textit{Right:} Corrected scaling factor $A/\sin(\alpha)$ extracted from voltage as a function of the electromagnetic energy in simulations. A first-order linear dependence is observed, with a small spread.  }
   \label{fig:simus_reconstruction}
\end{figure}

The angular resolution $\psi$, defined as the angle between the true shower direction and the one reconstructed using the process presented in Section~\ref{sec:recons}, is shown in Figure~\ref{fig:simus_reconstruction}. We find a median resolution of $0.09^\circ$. These results are comparable to those obtained using the electric field directly.

In the standard ADF treatment, the quantity $A/\sin(\alpha)$—where $A$ is the ADF scaling factor—has been shown to correlate linearly with the electromagnetic energy $E_{\rm em}$ to first order \cite{Guelfand_2025}, with second-order effects discussed in Section~\ref{sec:recons}. Figure~\ref{fig:simus_reconstruction} ({\it Right}) presents the equivalent distribution obtained from voltage-derived scaling factors. 
A clear first-order linear correlation between $A/\sin(\alpha)$ and $E_{\rm em}$ is observed.

While we do not compute a full energy estimator here, this result demonstrates that the quantity $A/\sin(\alpha)$ derived from voltage can serve as a robust proxy for estimating the electromagnetic energy of the shower.

These results validate the hypothesis that the ADF model can be applied successfully to voltage traces, enabling event selection (via $\chi^2_\nu$), direction reconstruction, and a first-order energy estimation—all without requiring antenna response deconvolution and electric field reconstruction.

\vspace{-0.3cm}
\section{Reconstruction of GRAND cosmic-ray candidates}
\vspace{-0.3cm}
\subsection{Commissioning data from GP300: context}
\vspace{-0.2cm}
The GP300 detector is currently in its commissioning phase, with 65 out of the planned 289 antennas deployed as of January 2025. During this phase, extensive efforts are being carried out on various aspects, including thermal regulation, suppression of radio self-emission, and detector calibration. In particular, the antenna-level trigger system (described in~\cite{Ma_ICRC2025}) is undergoing testing and optimization. This trigger uses six parameters—such as a minimum amplitude threshold—to identify radio pulses within the ambient background, which are recorded as Unit Data (UD).

Since December 2024, when at least five antennas trigger within a coincidence window (set to 10 $\mu \rm{s}$), the data are stored as Coincident Data (CD). Between December 2024 and March 2025, these CD events were analyzed by the GRAND Collaboration, leading to the selection of 41 cosmic-ray candidates, as detailed in \cite{Lavoisier_ICRC2025}.

In this work, we apply the ADF reconstruction method to these candidate events. We first use the voltage traces to refine the initial selection, and then apply the method to the reconstructed electric field signals to obtain a more accurate energy estimate.

\vspace{-0.2cm}
\subsection{Candidate selection with ADF}
\vspace{-0.2cm}

The ADF fit was applied on random CD events, without preselection —thus mostly background, though some cosmic-ray events may be present. Their $\chi^2_\nu$ distribution  (Figure~\ref{fig:voltage_CR0} {\it Left}) is clearly separated from the one associated with cosmic rays from our simulation set, with a 16\% fraction only passing the cut $\chi^2_\nu \leq 25$.
A detailed analysis showed that, in the vast majority of cases, these selected background events exhibit similar amplitudes on all antennas, with reconstructed angular distances $\omega$ close to the Cherenkov angle. It is expected that a larger detector, with optimal trigger efficiency, will allow for a better background rejection, as background sources then will trigger more antennas on a wider $\omega$ range.

To study further the potential of the amplitude pattern for background rejection,  we defined a very simple amplitude fit $A=\kappa/l$  --where $l$ is the distance from the antenna to the source and $\kappa$ an adjustable parameter-- corresponding to an istropic emission from the point source $X_e$. This background fit was applied to both simulated (see section \ref{subsec:recons_simus}) and CD data. Figure \ref{fig:voltage_CR0} shows that the 2 populations follow quiet distinct trends. Once the detector is fully commissioned and optimized, a combined cut on $\chi^2_\nu(\textrm{ADF})$ and $\chi^2_\nu(\textrm{bckgd})$ could allow for a powerful rejection of background, even more when the detector is extended to its final 289-units configuration.

\vspace{-0.2cm}
\subsection{ADF fit applied to candidate}
\begin{figure}[htbp]
   \centering
   \begin{minipage}[t]{0.4\linewidth}
      \centering
      \includegraphics[width=\linewidth]{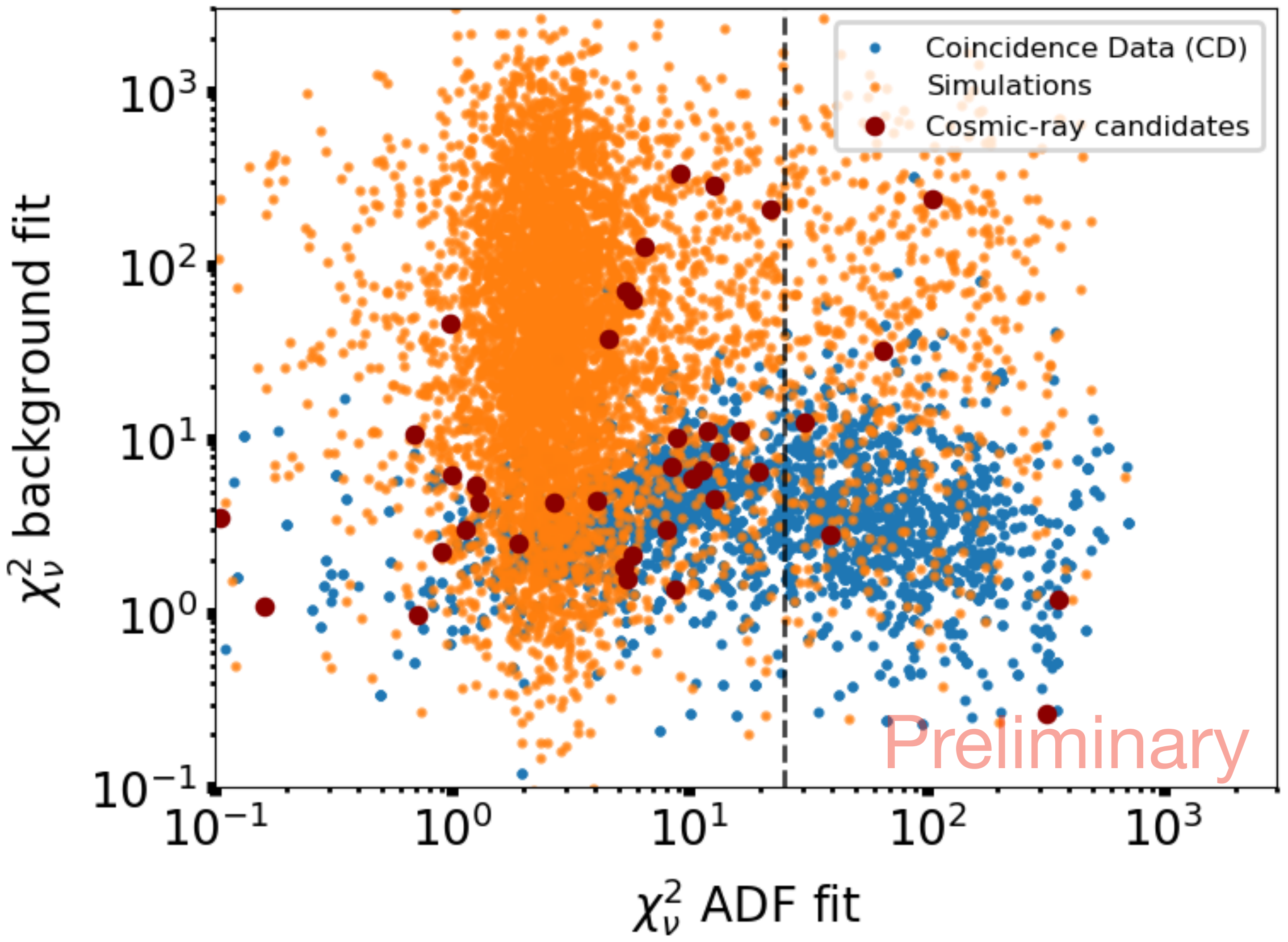}
   \end{minipage}
   \begin{minipage}[t]{0.4\linewidth}
      \centering
      \includegraphics[width=\linewidth]{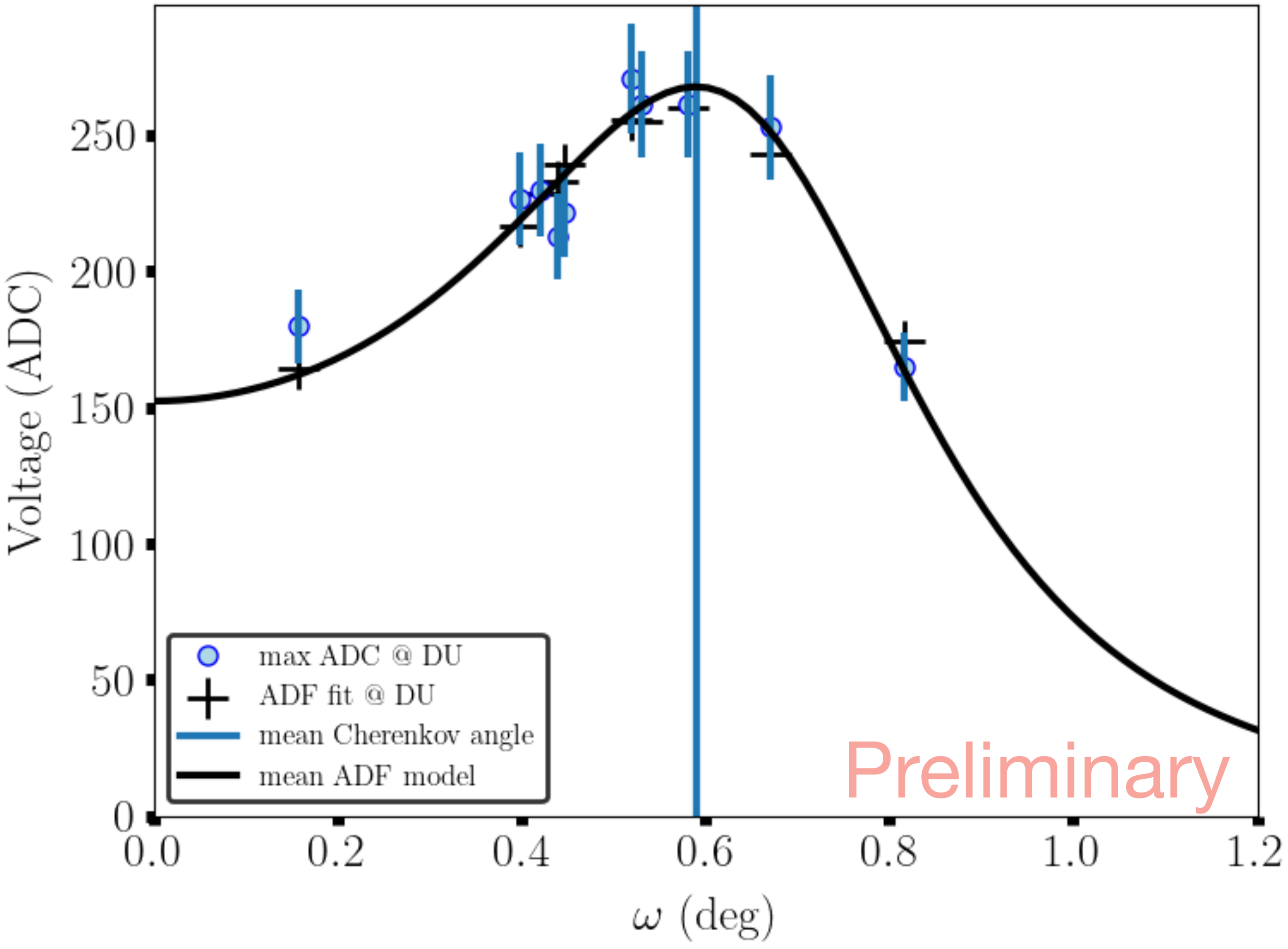}
   \end{minipage}

   \caption{{\it Left:} Distribution of the $\chi^2_\nu$ from the ADF and background fits for cosmic-ray simulations, random CD (interpreted as background events), and cosmic-ray candidates. A clear separation is observed between background and cosmic-ray events. Most CR candidates fall within the cosmic-ray distribution, while those with $\chi^2_\nu > 25$ are excluded from further reconstruction, as they are considered likely background.
   {\it Right:} Amplitude profile of one of the cosmic-ray candidates (CR0: $\theta=78.3\degree$, $\phi=137.8\degree$) with the ADF fit, with $\chi^2_\nu \sim 1$. Blue points show the measured peak amplitudes at each antenna as a function of their angular coordinate, with error bars representing the estimated 7.5\% uncertainties on the amplitude. The vertical blue line indicates the Cherenkov angle predicted by the toy model, at which the expected amplitude enhancement occurs. Black crosses correspond to the ADF fit at each antenna position and the black line represents the averaged ADF model across all $\eta$ directions.
   }
   \label{fig:voltage_CR0}
\end{figure}

\vspace{-0.2cm}
Finally the ADF fit was applied directly to the cosmic-ray candidate. As in simulations, only events with a reduced ADF chi-squared $\chi^2_\nu \leq 25$ were retained, representing 85\% of the sample (see Figure~\ref{fig:voltage_CR0}, {\it Left}), comparable with the fraction obtained for simulated events.

Figure~\ref{fig:voltage_CR0} ({\it Right}) illustrates an example of well-reconstructed candidate where the fitted profile (black line) closely matches the measured amplitudes (blue dots) from triggered antennas. A clear enhancement at the expected Cherenkov angle is observed, consistent with previous studies \cite{Corstanje_2017, 2011_deVries, 2010_Alvarez-Muniz}. The agreement between the predicted Cherenkov angle (which is a fixed parameter of the ADF fit) and the observed enhancement supports the cosmic-ray origin. 

\begin{figure}[htbp]
   \centering
   \begin{minipage}[t]{0.44\linewidth}
      \centering
      \includegraphics[width=\linewidth]{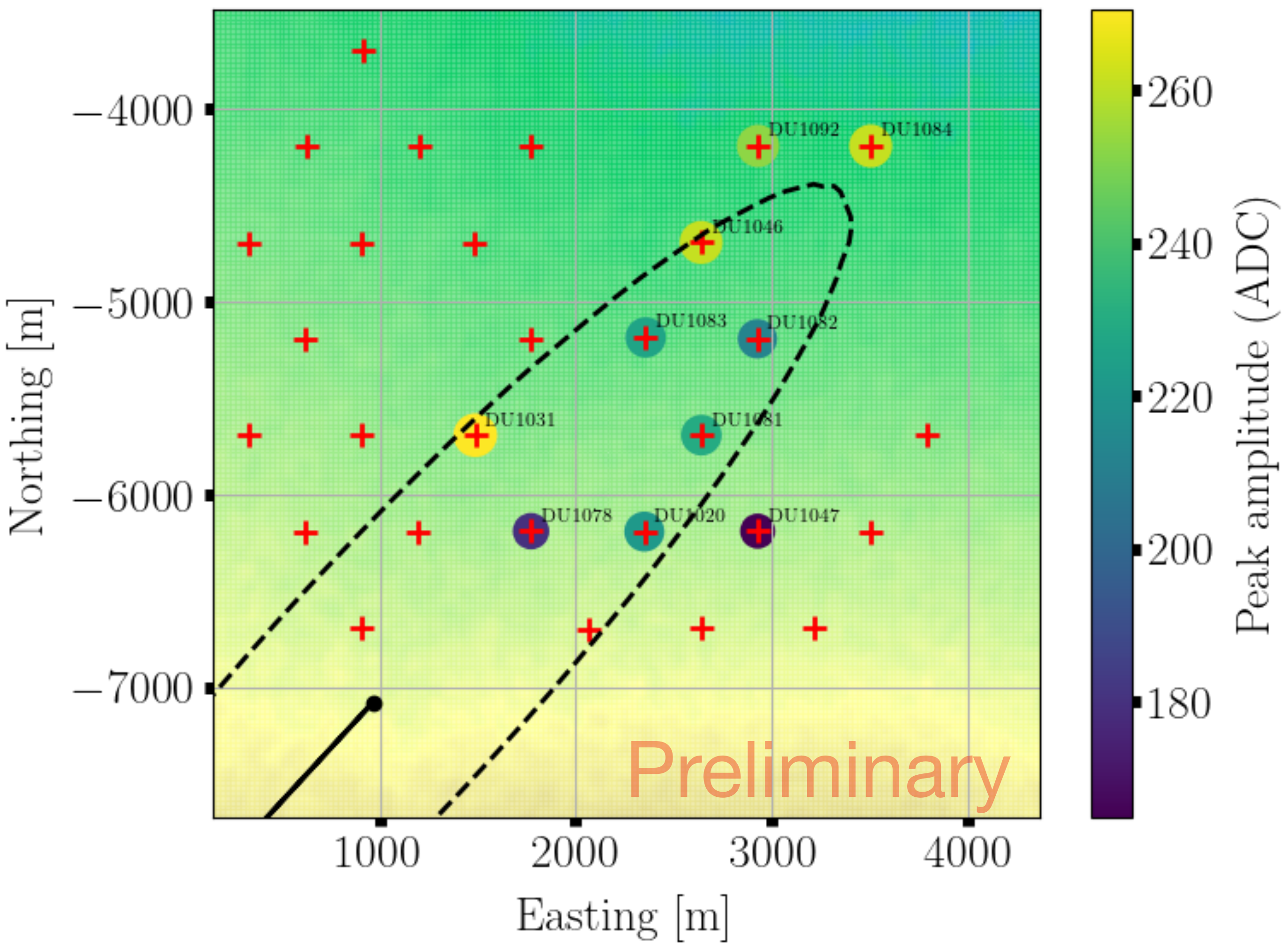}
   \end{minipage}
   \hfill
   \begin{minipage}[t]{0.48\linewidth}
      \centering
      \includegraphics[width=\linewidth]{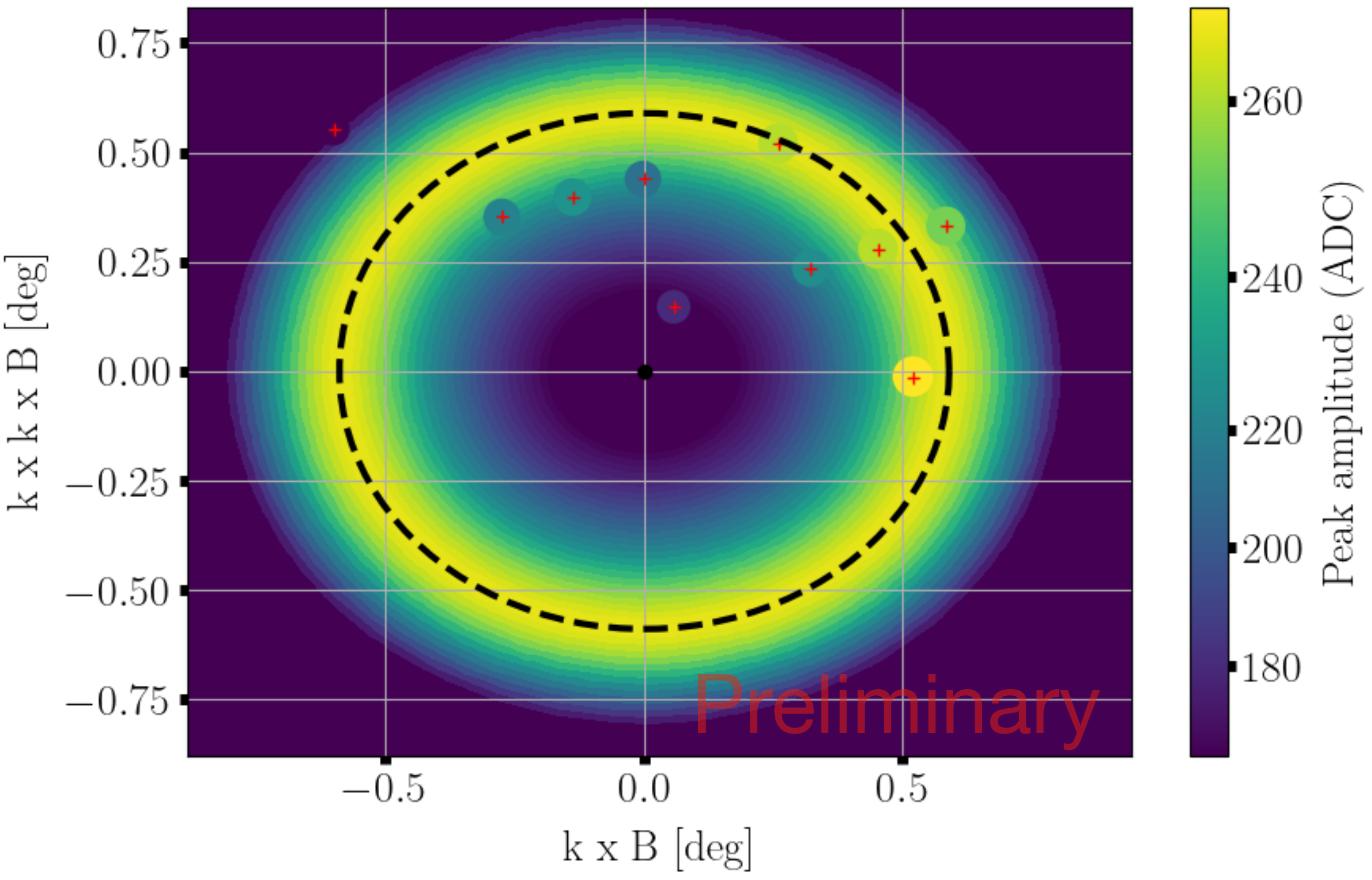}
   \end{minipage}
   \caption{
   {\it Left:} Map of active antennas on site for the cosmic-ray candidate (CR0), with amplitudes shown at triggered antennas. The reconstructed radio footprint is displayed.
   {\it Right:} Triggered antennas projected onto the reconstructed shower plane.}
   \label{fig:antenna_CR0}
\end{figure}

Figure~\ref{fig:antenna_CR0} ({\it Left}) shows the position of active antennas on site during the relevant data acquisition period for the same candidate, with circles indicating the peak amplitudes of triggered antennas. The reconstructed footprint --derived from the Cherenkov angle (from the toy model), the core position, and the shower axis-- is also displayed. The pattern is overall consistent, though a few antennas within the expected footprint did not trigger, likely due to the trigger system not yet being fully optimized.  
Figure~\ref{fig:antenna_CR0} ({\it Right}) shows the positions of the triggered antennas projected onto the reconstructed shower plane.

\begin{figure}[!htp]
   \centering
   \begin{minipage}[t]{0.4\linewidth} 
      \centering
      \includegraphics[width=\linewidth, height=5cm, keepaspectratio]{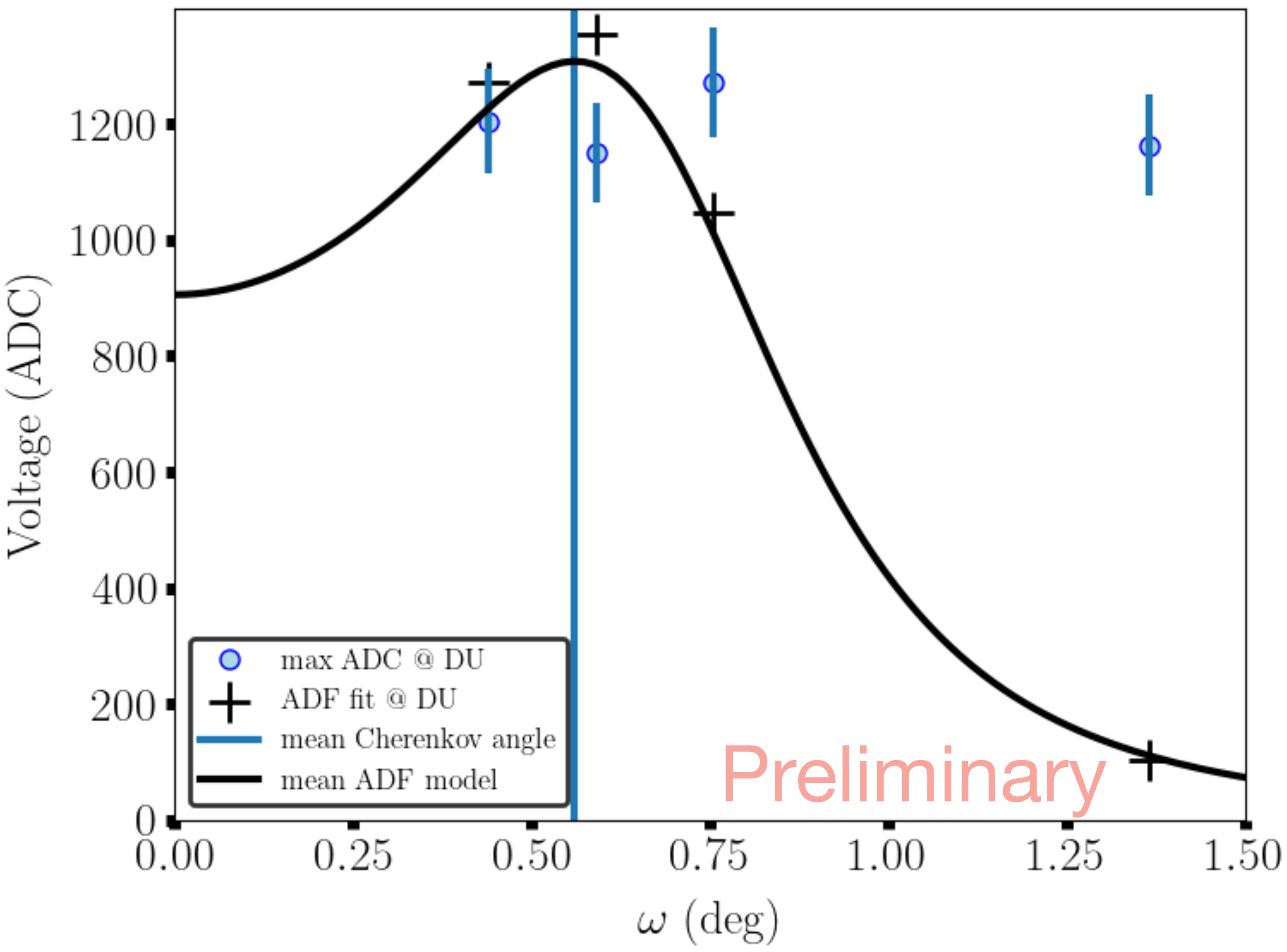} 
   \end{minipage}
   \begin{minipage}[t]{0.4\linewidth}
      \centering
      \includegraphics[width=\linewidth, height=5cm, keepaspectratio]{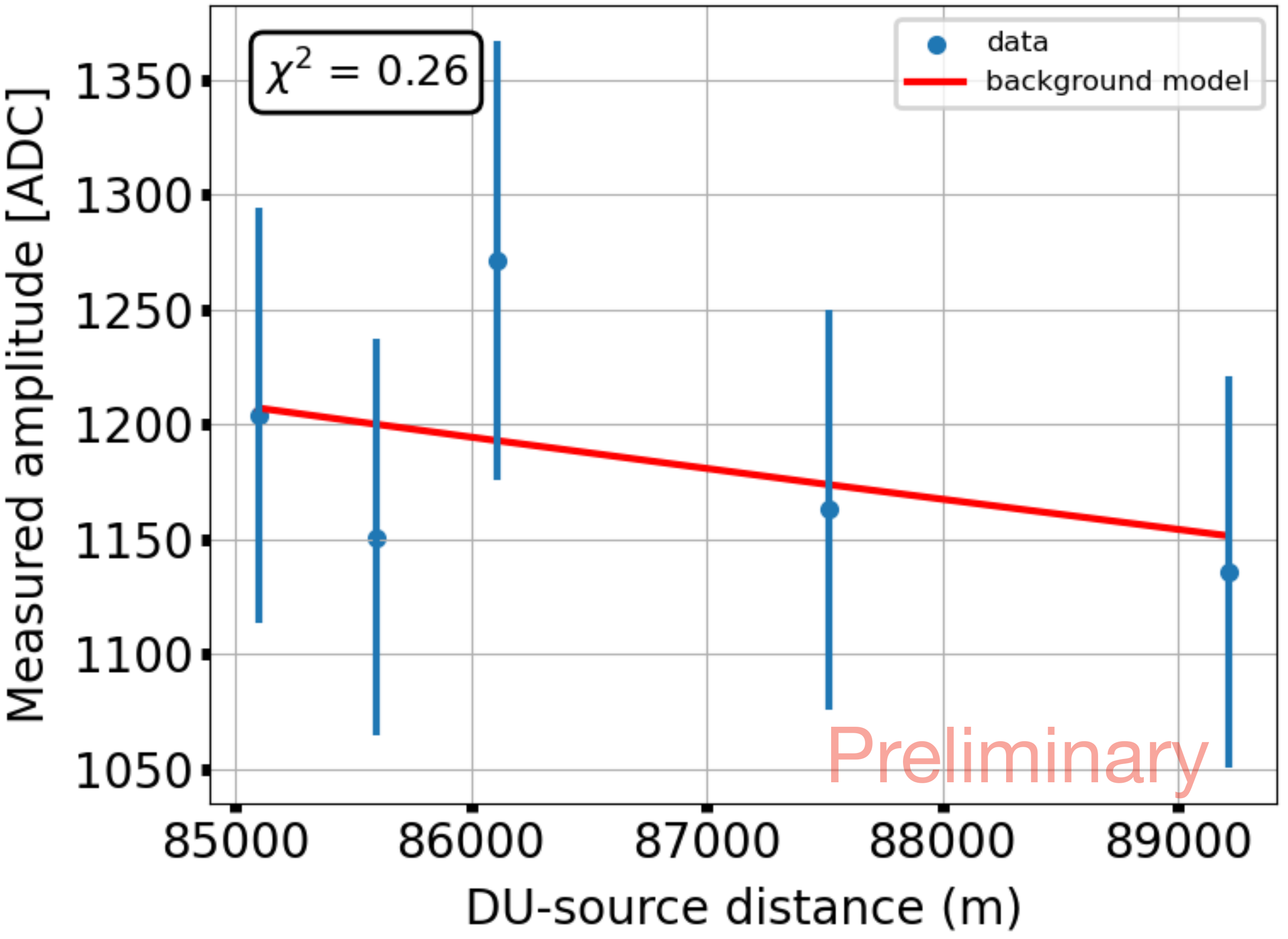}
   \end{minipage}
   \caption{{\it Left:} Amplitude profile of a cosmic-ray candidate (CR16: $\theta=80.8\degree$, $\phi=178.5\degree$) fitted with the ADF model, which does not meet the selection criterion $\chi^2_\nu \leq 25$.
    {\it Right:} Amplitude profile of the same event using the background fit. Blue points represent the amplitudes at triggered antennas, plotted as a function of their distance $l$ to the reconstructed source, with error bars representing the same estimated 7.5\% uncertainties on the amplitude.}
   \label{fig:voltage_CR16}
\end{figure}

In contrast, Figure~\ref{fig:voltage_CR16} ({\it Left}) shows an example of a candidate with  $\chi^2_\nu =321$ from the ADF fit, falling within the background distribution and thus rejected.

These results validate the ADF fit’s robustness on experimental voltage data, with reliable direction reconstruction and consistent footprints. This sets the stage for the use of ADF in energy reconstruction.

\vspace{-0.2cm}
\subsection{Reconstruction on Voltage and Electric Field}
\begin{figure}[htbp]
   \centering
   \begin{minipage}[t]{0.4\linewidth}
      \centering
      \includegraphics[width=\linewidth]{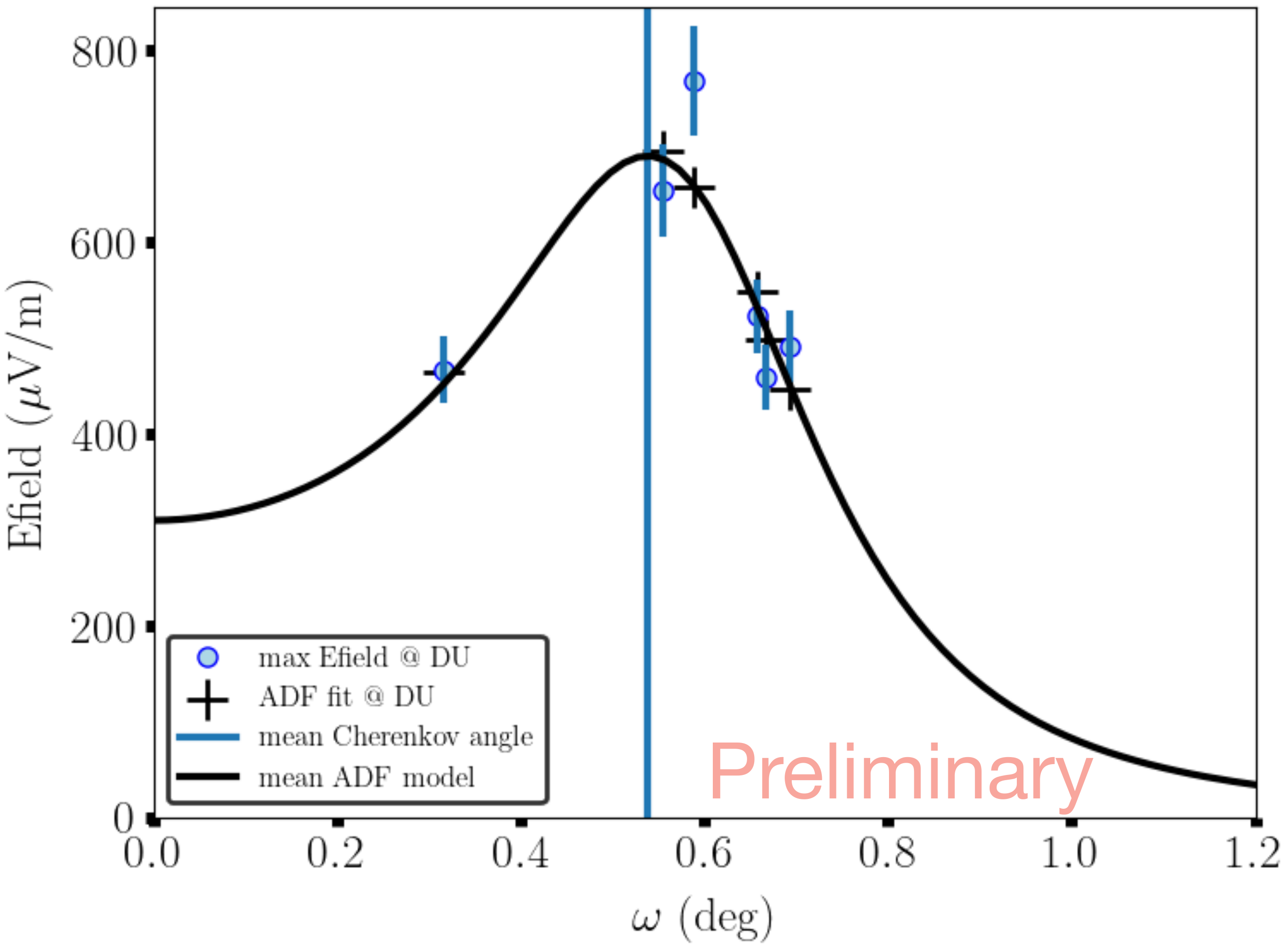}

   \end{minipage}
   \begin{minipage}[t]{0.4\linewidth}
      \centering
      \includegraphics[width=\linewidth]{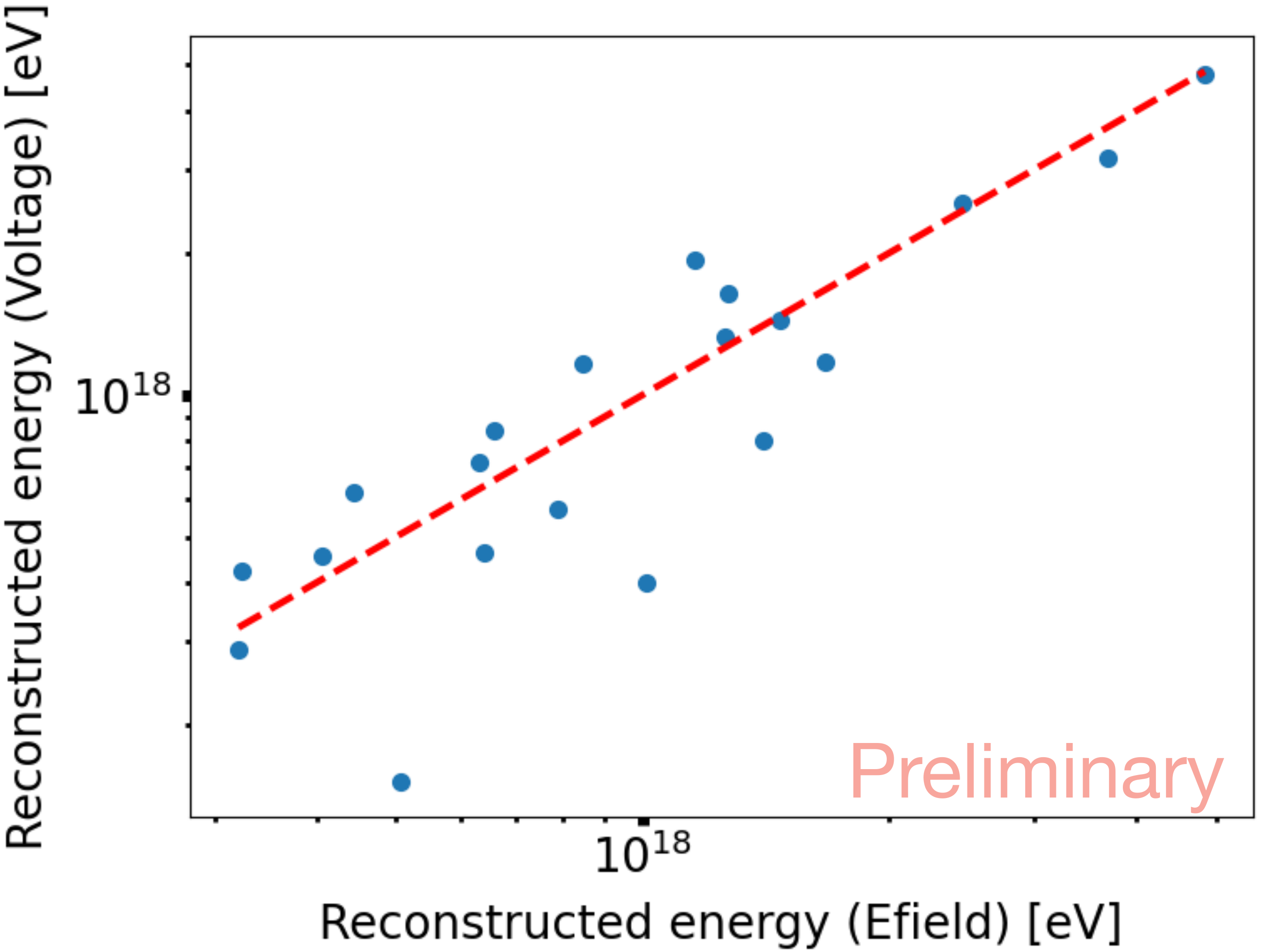}
   \end{minipage}
   \vspace{1em} 

   \begin{minipage}[t]{0.4\linewidth}
      \centering
            \includegraphics[width=\linewidth]{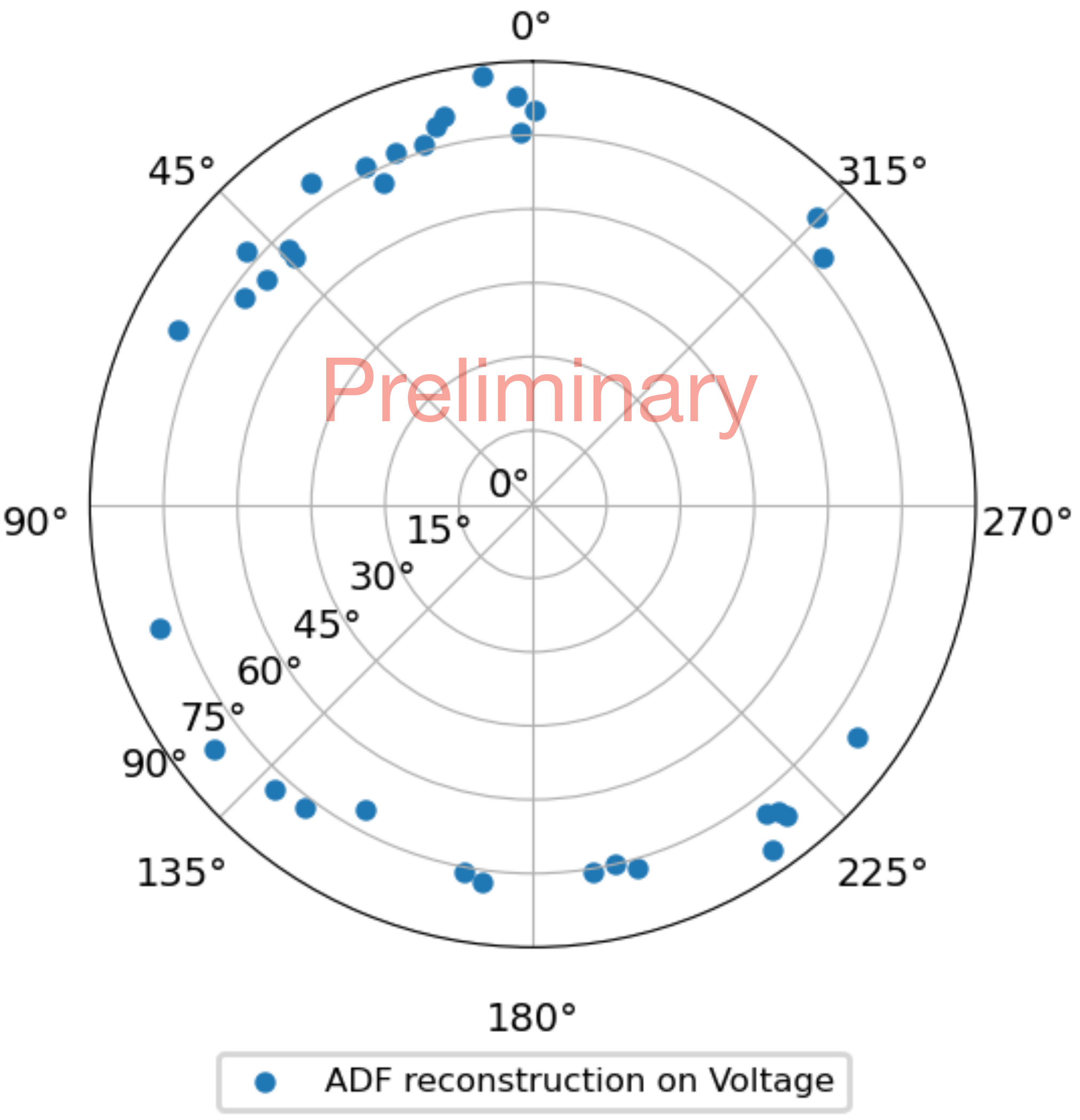}
   \end{minipage}

   \caption{{\it Top Left:} ADF fit applied to the reconstructed electric field for a selected candidate  (CR20: $\theta=77.41\degree$, $\phi=310.35\degree$).
   {\it Bottom:} Reconstructed arrival directions (zenith $\theta$ and azimuth $\phi$) of the cosmic-ray candidates from voltages. The zenith is at the center, the horizon at the outer circle, and $\phi$ is measured from North ($0^\circ$).
   {\it Top Right:} Correlation between the electromagnetic energy reconstructed from voltage signals and that from the reconstructed electric field for the cosmic-ray candidates. A small spread is observed, indicating a good agreement and supporting the method’s reliability.}
   
   \label{fig:recons_candidates}
\end{figure}

\vspace{-0.2cm}
As shown in Section~\ref{subsec:recons_simus}, ADF treatment of voltages offers excellent angular resolution in simulations. Figure~\ref{fig:recons_candidates} ({\it Bottom}) displays the reconstructed direction from voltages of the selected cosmic-ray candidates. 

The electric field is then reconstructed using the deconvolution procedure described in~\cite{zhang_2025}, with an example of an ADF fit on electric-field traces shown in Figure~\ref{fig:recons_candidates} ({\it Top Left}). The electromagnetic energy of each event is then estimated from this fit with the method described in Section~\ref{sec:recons}.

As a cross-check, the energy is also reconstructed directly from the voltage traces using the empirical scaling law derived from simulations (Section~\ref{subsec:recons_simus}). Figure~\ref{fig:recons_candidates} ({\it Top Right}) compares the reconstructed energies from both methods. A good agreement is observed, with a relative resolution (defined as $(E^*_{\rm em, voltage} - E^*_{\rm em, efield})/E^*_{\rm em, efield}$, where $E^*_{\rm em, efield}$ is the reconstructed energy from the reconstructed electric field and $E^*_{\rm em, voltage}$ is the reconstructed energy from voltage) of about 32\%, indicating that voltage-based reconstruction provides a reliable first-order energy estimate.

Moreover, the reconstructed energy distribution spans from $10^{17}$ eV to $5 \times 10^{18}$ eV, matching expectations based on our exposure calculations (see \cite{Kato_ICRC2025}), which further confirms the method’s robustness. It should be noted that the amplitude calibration has not yet been applied to either voltage traces or electric field reconstructions, so these results remain preliminary.

\vspace{-0.3cm}
\section{Conclusion}
\vspace{-0.3cm}
In this work, we have demonstrated the successful application of the ADF method directly to voltage signals recorded by the GP300 detector. By bypassing the full electric field reconstruction, this approach enables efficient cosmic-ray identification, accurate direction reconstruction, and a first-order estimation of the shower energy from raw voltage data.

The study underlines the potential of voltage-based reconstruction for large-scale radio arrays like GRAND.
Future work will focus on including  amplitude profile studies into the cosmic-ray candidate selection process,  refining the energy calibration and extending this technique to larger datasets as GP300 continues its commissioning phase.

\vspace{-0.3cm}
\bibliographystyle{ICRC}
\setlength{\bibsep}{0pt plus 0.3ex}
{\footnotesize
}
\bibliography{references}

\clearpage

\section*{Full Author List: GRAND Collaboration}

\scriptsize
\noindent
J.~Álvarez-Muñiz$^{1}$, R.~Alves Batista$^{2, 3}$, A.~Benoit-Lévy$^{4}$, T.~Bister$^{5, 6}$, M.~Bohacova$^{7}$, M.~Bustamante$^{8}$, W.~Carvalho$^{9}$, Y.~Chen$^{10, 11}$, L.~Cheng$^{12}$, S.~Chiche$^{13}$
% ...
\\
\\
$^{1}$Departamento de Física de Particulas \& Instituto Galego de Física de Altas Enerxías, Universidad de Santiago de Compostela, 15782 Santiago de Compostela, Spain \\
$^{2}$Institut d'Astrophysique de Paris, CNRS  UMR 7095, Sorbonne Université, 98 bis bd Arago 75014, Paris, France \\
$^{3}$Sorbonne Université, Université Paris Diderot, Sorbonne Paris Cité, CNRS, Laboratoire de Physique  Nucléaire et de Hautes Energies (LPNHE), 4 Place Jussieu, F-75252, Paris Cedex 5, France \\
$^{4}$Université Paris-Saclay, CEA, List,  F-91120 Palaiseau, France \\
$^{5}$Institute for Mathematics, Astrophysics and Particle Physics, Radboud Universiteit, Nijmegen, the Netherlands \\
$^{6}$Nikhef, National Institute for Subatomic Physics, Amsterdam, the Netherlands \\
$^{7}$Institute of Physics of the Czech Academy of Sciences, Na Slovance 1999/2, 182 00 Prague 8, Czechia \\
$^{8}$Niels Bohr International Academy, Niels Bohr Institute, University of Copenhagen, 2100 Copenhagen, Denmark \\
$^{9}$Faculty of Physics, University of Warsaw, Pasteura 5, 02-093 Warsaw, Poland \\
$^{10}$Key Laboratory of Dark Matter and Space Astronomy, Purple Mountain Observatory, Chinese Academy of Sciences, 210023 Nanjing, Jiangsu, China \\
$^{11}$School of Astronomy and Space Science, University of Science and Technology of China, 230026 Hefei Anhui, China \\
$^{12}$National Astronomical Observatories, Chinese Academy of Sciences, Beijing 100101, China \\
$^{13}$Inter-University Institute For High Energies (IIHE), Université libre de Bruxelles (ULB), Boulevard du Triomphe 2, 1050 Brussels, Belgium \\
% ...


\subsection*{Acknowledgments}

\noindent
The GRAND Collaboration is grateful to the local government of Dunhuag during site survey and deployment approval, to Tang Yu for his help on-site at the GRANDProto300 site, and to the Pierre Auger Collaboration, in particular, to the staff in Malarg\"ue, for the warm welcome and continuing support.
The GRAND Collaboration acknowledges the support from the following funding agencies and grants.
\textbf{Brazil}: Conselho Nacional de Desenvolvimento Cienti\'ifico e Tecnol\'ogico (CNPq); Funda\c{c}ão de Amparo \`a Pesquisa do Estado de Rio de Janeiro (FAPERJ); Coordena\c{c}ão Aperfei\c{c}oamento de Pessoal de N\'ivel Superior (CAPES).
\textbf{China}: National Natural Science Foundation (grant no.~12273114); NAOC, National SKA Program of China (grant no.~2020SKA0110200); Project for Young Scientists in Basic Research of Chinese Academy of Sciences (no.~YSBR-061); Program for Innovative Talents and Entrepreneurs in Jiangsu, and High-end Foreign Expert Introduction Program in China (no.~G2023061006L); China Scholarship Council (no.~202306010363); and special funding from Purple Mountain Observatory.
\textbf{Denmark}: Villum Fonden (project no.~29388).
\textbf{France}: ``Emergences'' Programme of Sorbonne Universit\'e; France-China Particle Physics Laboratory; Programme National des Hautes Energies of INSU; for IAP---Agence Nationale de la Recherche (``APACHE'' ANR-16-CE31-0001, ``NUTRIG'' ANR-21-CE31-0025, ANR-23-CPJ1-0103-01), CNRS Programme IEA Argentine (``ASTRONU'', 303475), CNRS Programme Blanc MITI (``GRAND'' 2023.1 268448), CNRS Programme AMORCE (``GRAND'' 258540); Fulbright-France Programme; IAP+LPNHE---Programme National des Hautes Energies of CNRS/INSU with INP and IN2P3, co-funded by CEA and CNES; IAP+LPNHE+KIT---NuTRIG project, Agence Nationale de la Recherche (ANR-21-CE31-0025); IAP+VUB: PHC TOURNESOL programme 48705Z. 
\textbf{Germany}: NuTRIG project, Deutsche Forschungsgemeinschaft (DFG, Projektnummer 490843803); Helmholtz—OCPC Postdoc-Program.
\textbf{Poland}: Polish National Agency for Academic Exchange within Polish Returns Program no.~PPN/PPO/2020/1/00024/U/00001,174; National Science Centre Poland for NCN OPUS grant no.~2022/45/B/ST2/0288.
\textbf{USA}: U.S. National Science Foundation under Grant No.~2418730.
Computer simulations were performed using computing resources at the CCIN2P3 Computing Centre (Lyon/Villeurbanne, France), partnership between CNRS/IN2P3 and CEA/DSM/Irfu, and computing resources supported by the Chinese Academy of Sciences.

\end{document}